\documentstyle[12pt,epsf]{article}
\newcommand\be{\begin{equation}}
\newcommand\ee{\end{equation}}
\newcommand\ba{\begin{eqnarray}}
\newcommand\ea{\end{eqnarray}}
\newcommand\tr{{\rm tr}}
\begin{document}

\title{\bf Thermodynamics of Constituent Quarks\thanks{Supported 
by 
the
 Bundesministerium f\"ur Forschung und Technologie (BMBF) under
grant no. 06 HD 742 }}

\author{H.J. Pirner and M. Wachs\\\\
Institut f\"ur Theoretische Physik,\\ 
Universit\"at
Heidelberg}
\date{ }

\vspace{1.0cm}

\setlength{\baselineskip}{24pt}

\maketitle

\begin{abstract}
\setlength{\baselineskip}{18pt}
\noindent
We investigate the thermal behavior of quarks and antiquarks 
interacting
via a temperature-dependent linear potential. The quarks are 
constituent
quarks with dynamically generated masses from the background 
linear $\sigma$-model.
\end{abstract}


\section{Introduction}

The constituent quark model has been rather successful in meson 
and
baryon spectroscopy \cite{LSG}. Its main ingredients are massive 
$(m\simeq 
300$ MeV) quark constituents interacting with a potential, which is 
linear at large
distances and coulombic at short distances. A smaller but important 
role is played by
spin-dependent  exchange interactions which modify the 
mass spectra.
Recent work on the constituent quark model emphasizes the 
dynamical nature of the
constituent quark mass. A chiral $\sigma$-model 
\cite {G} may be used to
generate the quark mass thereby allowing the  important coupling 
of the constituent
quark to its pion cloud. By analysing electron scattering
on the proton an estimate \cite {BPKP}
of the pion cloud of the constituent quark has been obtained. 
In ref. \cite {GR} its  effects
on baryon spectroscopy have been worked out. The linear $\sigma$-model 
coupled to quarks
bears some resemblances to the very popular Nambu-Jona-Lasinio 
model \cite {HK,K}, but it
treats the $0^+$ and $0^-$ mesons as fundamental fields. This is an 
assumption which
is true if the system is analysed with a low resolution 
probe or at
moderate temperatures. We would like to argue that up to a
temperature $T\approx 250$,
MeV i.e. beyond the chiral transition temperature 
it is still reasonable to separate $(\sigma,\vec\pi)$ degrees of 
freedom from the rest of other mesons.
Indications in
this direction come from lattice simulations \cite {BG}. 
Treating  the thermodynamics of the constituent quarks, we
allow a deconfinement
transition. The thermal gluon fluctuations are modeled to lead
to a
decreasing string
tension with temperature. We neglect all feedback from 
the quark dynamics
to the gluon dynamics. Therefore the decrease of the string tension 
is given in our
model from pure gluon QCD. This way the chiral transition 
tem\-pe\-ra\-ture may differ
from the temperature where the string tension goes to zero. 
Actual lattice simulations  show a coincidence of the strongly
decreasing value of the quark condensate
$\langle\bar q q\rangle$ and an increasing expectation value of the 
Wilson loop
operator. For dynamical
quarks the Polyakov line operator, 
however, is no longer an order parameter.
The problem of
double counting the scalar and pseudoscalar degrees of freedom 
 $(\sigma,\vec\pi)$ as $\bar q q$ bound states and
as explicit degrees of freedom can be solved by
excluding mesons with  quantum numbers $(^1 S_0$ and $^3 P_0$)
in the summation over bound $q\bar q$ states.

The partition function of the constituent quark model is given as:
\be{\cal Z}=\int{\cal D}\psi{\cal D}{\bar\psi}{\cal D}\phi
e^{-\int^\beta_0 d\tau\int d^3 x{\cal L}(\psi,\bar\psi,\phi)}\ee
with the Euclidean Lagrangian
\begin{eqnarray}
{\cal L}&=&\bar\psi\left(i\gamma\partial-
g(\sigma+i\vec\pi\vec\tau
\gamma_5 )\right)\psi+(\bar\psi\Gamma\psi)_x
V(x-y)(\bar\psi\Gamma\psi)_y\nonumber\\
&&+\frac{1}{2}\partial_\mu\vec\phi\partial_\mu\vec\phi+
U(\vec\phi),
\end{eqnarray}
where 
$U(\vec\phi)=-
\frac{\mu^2_0}{2}\vec\phi\vec\phi+\frac{\lambda}{4}(\vec\phi\vec\phi)^2$.
\\
Here $\psi=\left(\begin{array}{c} u\\d\end{array}\right)$ 
is the
 spinor of $u$ and $d$ quarks. The meson fields are combined 
into the four-component vector
\be\vec\phi=(\sigma,\vec\pi).
\ee
The parameters of the linear $\sigma$-model
at $T=0$ are chosen in such a way
that the minimum of the potential
$U(\vec\phi)$
lies at  $\langle\sigma_0\rangle=0.093$ GeV.
The light constituent quarks have a mass of
$m=300$ MeV and the $\sigma$ mass $m_{\sigma}=2 m$.
Then the coup\-lings  are $\mu^2_0=(0.495$ GeV$)^2,
\lambda=28.33, g=3.23$. These couplings are taken as 
temperature-independent.
%
The potential $V$ between $q\bar q$ in a 
color singlet state is
given by $V_{q\bar q}(\vec x-\vec y)=\kappa|\vec x-\vec y|-
2\sqrt{\kappa}$ with the
string tension $\kappa=\kappa_{0}=(447.2 $MeV)$^2$
at zero temperature. The second 
term takes into account
the self energy correction in the constituent quark model \cite{LSG}.
We do not
explicitly consider the gluon dynamics, but  we consider the string 
tension 
as a function of
temperature. 
Lattice simulations for $SU3_{color}$ \cite {KA,BEKL} 
have determined the critical temperature
$T_c=260$ MeV but not yet the critical behavior, how
the
string tension decreases with temperature.
For large space time dimensions $d$
it is possible to expand around a nontrivial stationary point of the 
string action. In a $1/d$ -expansion the decrease of the string tension 
has been calculated as a function of temperature \cite {PA}. 
The effective string tension at finite temperature is
\be \kappa_{eff}=\kappa_{0} (1 - (\frac {T}{T_c})^2 )^{\frac {1}{2}}\ee
with 
\be \frac {T_c^2}{\kappa }= \frac{3}{\pi d} + O (d^{-2}).\ee
Putting $d=4$ and neglecting the higher order corrections would
give a transition temperature of $220$ MeV.  We choose the
value $T_c=260$ MeV from lattice simulations. The decrease of $\kappa_{eff}$
at small temperatures is analogous to the universal Coulomb correction
in addition to the linear potential $\kappa r$.

The above simplified partition function  is still rather 
hard to solve, since the 
bound states
formed at low temperatures will grow in size and overlap with
increasing temperature. The 
resulting problem of a
color correlated system of overlapping quark antiquark clusters can 
probably only be
solved with a Monte Carlo calculation \cite {W}. We will limit 
ourselves in this
paper to a simplified analytical calculation, where in the low 
temperature limit
we evaluate the $q\bar q$
partition function  as  a sum over meson bound states. In the high 
temperature limit
we approximate  the quarks and
antiquarks as a correlated  gas with the string potential acting
as a perturbation.

The outline of the paper is as follows. In section 2 we calculate 
the spectrum of a single bound $q\bar q$ meson coupled to a heat 
bath,
in section 3 and 4 we compute the partition function of the many-quark 
and
antiquark system in the low and high temperature schemes given 
above.
In section 5 we address the chiral phase transition in mean field 
theory.

\section{Spectrum of a single $q\bar q$ bound state coupled to a 
heat bath}

 In a first step we neglect the
``elementary''  
mesons $(\sigma,
\vec\pi)$ and take all mesons as bound states of quarks and 
antiquarks with 
fixed quark masses. This approximation 
destroys the nice
low temperature behavior of the chiral $\sigma$-model, which is 
governed by the low
lying Goldstone pions, but we use  this 
simplification to show our new methods to solve the relativistic
bound state problem.  
This technical development is one 
of the  main new results
of this paper. The same method  
can also be included  in more evolved Monte Carlo 
calculations
at finite temperature and finite density. Note that the calculation has to 
be
relativistic since with increasing temperature the mean field mass 
of the quarks
will decrease, so we cannot use the framework of the 
nonrelativistic
constituent model. Instead we propose to treat the partition function of 
relativistic
quarks with the help of auxiliary variables. We start with the 
example of a single
meson composed of a quark and antiquark 
with total spin $S=0$ coupled to a heat bath. The quarks 
have fixed masses
$m$; then
\be Z_{q\bar q}=\tr\exp\left\{-\beta\left(\sqrt{\vec p_ 
q^2+m^2}+\sqrt{\vec p_{\bar
q}^2+m^2}+V(|\vec r_{\bar q}-\vec r_q|)\right)\right\}.\ee
The trick is to rewrite the relativistic Boltzmann factor as follows:
\be\exp(-\beta\sqrt{\vec 
p^2+m^2})=\frac{2}{\sqrt{\pi}}\int^\infty_0
d\mu\exp\left(-\mu^2-\frac{\beta^2}{4\mu^2}(\vec p^2+m^2)\right).\ee
In the appendix we outline the derivation of the full
partition function using the same parametrization of
the exponential integral as above.
Then $Z_{q\bar q}$ is given as 
\begin{eqnarray}
Z_{q\bar q}&=&\tr\frac{4}{\pi}\int^\infty_0 d\mu_1\int^\infty_0 
d\mu_2 e^{-\tilde{\cal H}}\nonumber\\
\tilde{\cal H}&=&\mu^2_1+\mu^2_2+\frac{\beta^2}{4\mu^2_1}(\vec
p^2_q+m^2)+\frac{\beta^2}{4\mu^2_2}(\vec p^2_{\bar q}+m^2)+\beta 
V(r).
\end{eqnarray}
This allows to separate relative and c.m. motion in the relativistic 
two-body
problem. Of course, we still have a static potential without 
retardation effects in the interaction. We also do not know how to 
include 
relativistic spin-spin  and spin-orbit interactions.

The relative and c.m. coordinates are:
\begin{eqnarray}
r=\vec r_q-\vec r_{\bar q}&&\qquad\qquad\vec 
p=\frac{\mu^2_2\vec p_q-\mu^2_1\vec
p_{\bar q}}{\mu^2_1+\mu^2_2}\nonumber\\
R=\frac{\mu^2_1\vec r_q+\mu^2_2\vec r_{\bar 
q}}{\mu^2_1+\mu^2_2}&&\qquad\qquad\vec
P=\vec p_q+\vec p_{\bar q}.\end{eqnarray}
With these coordinates the pseudo Hamiltonian $\tilde{\cal H}$ has 
the form
\begin{eqnarray}
\tilde{\cal H}&=&\mu^2_1+\mu^2_2+
\frac{\beta^2m^2}{4}\left(\frac{1}{\mu^2_1}+\frac{1}{\mu^2_2}
\right)+\frac{\beta^2}{4}\left[\frac{\vec 
P^2}{\mu^2_1+\mu^2_2}+\frac{\vec p^2}
{\mu^2_1\mu^2_2}(\mu^2_1+\mu^2_2)\right]\nonumber\\
&&+ \beta V(r).
\end{eqnarray}
At this stage it is advantageous to convert to new variables
\be x=\frac{\mu^2_1+\mu^2_2}{\sqrt2\beta}\qquad \mbox{ 
and}\qquad
y=\frac{\mu^2_2-\mu^2_1}{\sqrt2\beta}\ee
which give
\begin{eqnarray}
Z_{q\bar q}&=&\tr\frac{\sqrt2}{\pi}\beta\int^\infty_0 dx\int^x_{-x}
dy\frac{1}{\sqrt{x^2-y^2}}\exp(-\beta h)\nonumber\\
h&=&\sqrt2 x+\frac{m^2}{2^{3/2}(x-y)}+\frac{m^2}{2^{3/2}(x+y)}
+h_{\rm cm}+h_{\rm rel};\nonumber\\
h_{\rm cm}&=&\frac{\vec P^2}{2^{5/2}x}\nonumber\\
h_{\rm rel}&=&\frac{\vec p^2}{m_{\rm red}}+\kappa_{\rm eff}r
-2\sqrt{\kappa_{\rm eff}}.\nonumber\\
\end{eqnarray}
We see that the separation into a free c.m. Hamiltonian  $h_{\rm
cm}$ and a
Hamiltonian of relative motion $h_{\rm rel}$ has been achieved.
The c.m. kinetic energy contains an effective ``mass'' proportional to 
$x$,
whereas the energy of relative motion is proportional 
to the inverse reduced ``mass'' $m_{\rm red}$
\begin{eqnarray}
 m_{\rm red}&=&\sqrt2\left(x-\frac{y^2}{x}\right).
\end{eqnarray}
In 
the evaluation of
the trace we have to sum over all eigenstates of $h_{\rm cm}$ 
and $h_{\rm rel}$.
The eigenvalues of  $h_{\rm cm}$ are plane waves 
\be h_{\rm cm}|\vec K\rangle=\frac{\vec K^2}{2^{5/2}x}|\vec 
K\rangle.\ee
The eigenvalues of $h_{\rm rel}$ can be solved numerically:
\begin{eqnarray}
h_{\rm rel}\Psi_{n_r,\ell}(r)&=&\left(-\frac{\vec\nabla^2}{m_{\rm 
red}}+\kappa_{\rm eff}
r-2\sqrt{\kappa_{\rm eff}}\right)\Psi_{n_r,\ell}(
r)=\omega_{n_r,\ell}
\Psi_{n_r,\ell}(r)\nonumber\\
\omega_{n_r,\ell}&=&\alpha_{n_r,\ell}\sqrt[3]{\frac{\kappa^2_{\rm 
eff}}{m_{\rm
red}}}-2\sqrt{\kappa_{\rm eff}}.
\end{eqnarray}
For the lowest lying states the coefficients $\alpha_{n_r,\ell}$ are as 
follows:
$\alpha_{00}=2.34,\ \alpha_{01}=3.36,\ \alpha_{10}=4.09$ and 
$\alpha_{02}=4.25$.

To calculate the equation of state we need  the whole spectrum. 
Instead of numerically calculating the whole spectrum,  we 
take for 
the higher lying $\bar q q$ states the eigenstates of the harmonic 
oscillator 
with the
ground state adjusted to the exact solution. The high lying
states become important at high temperatures, where this approximation
is sufficient to get a qualitatively correct picture
\be 
\omega_{n_r,\ell}\approx(2n_r+\ell+2.34)\sqrt[3]{\frac{\kappa^2_
{\rm eff}}{m_{\rm
red}}}-2\sqrt{\kappa_{\rm eff}}.\ee
In principal this 
approximation is not  necessary, but it 
allows us to use analytical formulas in the following.
With the main quantum number
$n=2n_r+\ell$
the degeneracy of harmonic 
oscillator states
is $g(n)=(n+1)(\frac{n}{2}+1)$,
the energies are
$\omega(n)=(n+2.34)\sqrt[3]{\frac{\kappa^2_{\rm eff}}{m_{\rm
red}}}-2\sqrt{\kappa_{\rm eff}}$ and 
\be Z_{q\bar q}=\sum_{\vec K}\sum_n 
g(n)\frac{\sqrt{2}}{\pi}\int^\infty_0
dx\int^{+x}_{-x} dy \frac{ e^{-\beta\left[\Omega+\frac{\vec 
K^2}{2^{5/2}
x}\right]}}{\sqrt{x^2-y^2}}\ee
with
\be\Omega=\sqrt2 x+\frac{m^2_q}{2^{3/2}(x-y)}+\frac{m^2_{\bar 
q}}{2^{3/2}(x+y)}
+\omega (n).\ee
The c.m. motion can be integrated out giving the final partition 
function
\be Z_{q\bar
q}=\frac{2^{5/4}V}{\pi^{(5/2)}}\beta^{-0.5}\sum_n(n+1)
\left(\frac{n}{2}+1\right)\int^\infty_0 dx x^{1.5}\int^{+x}_{-x} dy 
\frac{1}
{\sqrt{x^2-y^2}}e^{-\beta\Omega}.\ee
We 
project out the bound state meson
masses at $T=0$:
\be\label{x} 
M(n_r,\ell)\bigm\vert_{T=0}=\lim_{\beta\to\infty}\left(-
\frac{1}{\beta}
\log
Z_{q\bar q}\right);\ee 
\begin{table}{{\bf Table 1:} Ground state masses $M$ from a
relativistic 
 and  nonrelativistic calculation for light 
$q\bar q$ and
strange quark $s\bar s$ states
(in brackets)
The experimental $\rho$-meson spectrum is given in addition.}\\[1ex]
\begin{tabular}{|l|c|c|c|c|}\hline
 $ M(n_r,\ell)$& $1s(0,0)$& $1p 
(0,1)$&$2s(1,0)$&
$2d(0,2)$\\\hline
relativistic (MeV)& $787\pm 7$ & $1205\pm14$ & $1494\pm 15$ 
& $1556\pm 15$\\
{}&$(1002\pm 7)$ & $(1391\pm15)$ & $(1665+15)$ & $(1724\pm 
16)$\\
nonrelativistic (MeV) & 900 & 1422.0 & 1793.1 & 1875\\
{}& (1064) & (1517) & (1838) & (2190)\\
exp (MeV) & 770 & 1260 & 1450 & 1700\\
{}&(1020) & (1285) & (1680) & {}\\
\hline
\end{tabular}
\end{table}

In table 1 we give the mass spectrum of relativistic quark 
antiquark states.
The bound states are calculated with fixed constituent masses
of $300$ MeV for the light quarks and $450$ MeV for the strange quarks. 
The numerical errors are estimated
in the following way: The lower values of the masses are obtained 
from eq. (\ref{x}) with only the specific
state in the partition function.
 The upper values are gotten by taking into account
finite temperature 
corrections using
the free meson partition function
\be\label{XX}
Z=\frac{V}{\lambda^3_T}e^{-\beta M}\ee
with $\lambda_T=(\frac{2\pi}{M(n_r,\ell) T})^{1/2}$ from
eq. (\ref{x})
for the 
particular state we are
interested in. 
To compute the
upper limits in
table 1 we apply eq. (\ref{XX}) with $Z$ = $Z_{q \bar q}$.
The minimal temperatures are 1.5 MeV for the ground state
and 3.0 MeV for
the excited states. 
The obtained spectrum  corresponds approximately to the $\rho$
meson 
spectrum with $\rho(1s)$, 
$\rho(2s)$, $\rho(2d)$ and the orbital $(\ell=1)$ excitation 
$a_1(1p)$. 
The agreement of the theoretical energies with experiment is rather 
good.
Especially the low lying 2s-state is improved by the relativistic
Hamiltonian. The splitting between the 2s- and 1s-state of 900 
MeV is lowered
to the experimentally observed 700 MeV.
Note, however, that the spin-dependent interactions are missing.
For a constituent quark mass of $450$ MeV we obtain the 
equivalent $\phi$-states.
Since the physical vector meson states are well described, we 
can explore the dependence of the meson masses on the choice 
of
quark masses and string tension. Both of these parameters vary 
with temperature.
In fig. 1 we give the dependence of the ground 
state energy on
the constituent mass and on the square root of the 
string
tension. In both cases the ground state energy drops. 
In the limit of small quark masses the meson mass is of 
order of $\sqrt\kappa_{0}$. 
For vanishing string tension the meson mass
converges towards twice the quark mass. Here the numerical error 
of $10$ MeV is comparable to the 
error estimates for the realistic meson masses given in table 1. 
\begin{figure}[hbt]
\unitlength1cm
\begin{center}
\begin{picture}(15,9)
\epsfbox{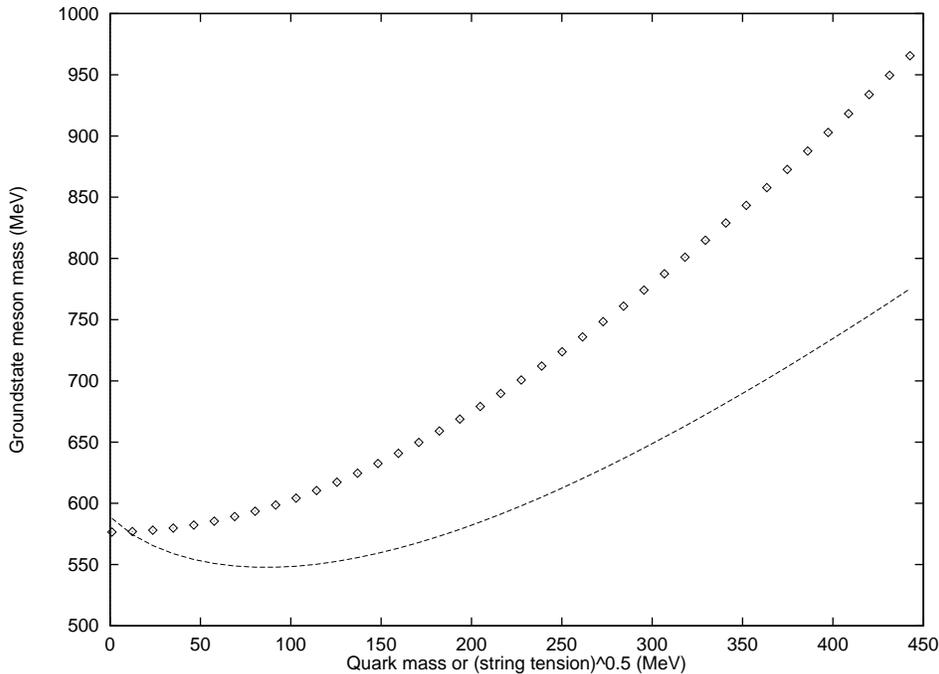}
\end{picture}
\end{center}
\caption{\label{afig1} Ground state meson mass  as a function
of constituent quark mass (squares) and as a function of
the square root of the string tension (dashed line).}
\end{figure}

\section{Nonoverlapping $(q\bar q)$-mesons at low temperature}

Now, we address the problem to calculate
the partition function $Z$ of a system of $N$-quarks and $N$-antiquarks. It 
can be 
written as
\ba 
Z&=&\tr\exp(-\beta{\cal H}_N)\nonumber\\
{\cal H}_N&=&\sum^N_{j=1}\sqrt{\vec 
p^2_{qi}+m^2}+\sum^N_{j=1}\sqrt{\vec p^2_{\bar
q j}+m^2}\nonumber\\
&&+\frac{1}{2}\sum^N_{i,j} V(\vec r_{qi}-\vec r_{\bar q j}).\ea
As before we are limited in our treatment to nonretarded 
interactions given by the
static string tension. In addition we have to
sum over all
possible arrangements of flux tubes connecting $q\bar q$ pairs. To 
our knowledge this
is only possible in a fully dynamical way with a Monte 
Carlo model, as
has been proposed by the Ontario group \cite {W}. In the following 
we limit 
ourselves to low temperatures,
where the number of mesons is small. Therefore we can safely assume that 
the individual 
mesons keep their
identity and do not interact. 
Consequently we 
have a
$N$-``meson'' partition function with the $i$'th quark always 
correlated with  the 
$i$'th antiquark
\be
Z=\tr\left(\frac{4}{\pi}\right)^N\prod^N_{i=1}\int^\infty_0 
d\mu_{1,i}\int^\infty_0
d\mu_{2,i}\exp-[{\cal H}(i)]\ee
where for the $i$'th meson ${\cal H}(i)$ is the same Hamiltonian as 
in the single
meson problem
\ba
{\cal H}(i)&=&\mu^2_{1,i}+\mu^2_{2,i}+\frac{\beta^2}{4\mu^2_{1,i}}
(\vec p_{qi}+m^2)\nonumber\\
&&+\frac{\beta^2}{4\mu^2_{2,i}}(\vec p^2_{\bar q,i}+m^2)
+\beta(\kappa_{\rm eff} r_i-2\sqrt\kappa_{\rm eff}).\ea
After the introduction of c.m. and relative coordinates we transform
to the
variables $x_i$ and $y_i$ as before and get
\be 
Z=\tr\left(\frac{\sqrt2}{\pi}\beta\right)^N\prod_{i=1}^N\int^\infty_
0 dx_i
\int^{+x_i}_{-x_i}\frac{dy_i}{\sqrt{x^2_i-y^2_i}}\exp[-\beta h(i)]\ee
with \be h(i)=\sqrt2 x_i+\frac{m^2}{2^{3/2}(x_i-
y_i)}+\frac{m^2}
{2^{3/2}(x_i+y_i)}+\hat h_{s_i}+\hat h_{{\rm rel},i}\ee
where 
\ba
\hat h_{s_i}&=&\frac{\vec P_i^2}{2^{5/2}x_i};\nonumber\\
\hat h_{{\rm rel},i}&=&\frac{\vec p^2_i}{m_{\rm 
red}(i)}+\kappa_{\rm eff}
r_i-2\sqrt\kappa_{\rm eff};\nonumber\\
m_{\rm red}(i)&=&\sqrt2\left(x_i-\frac{y^2_i}{x_i}\right).\ea
We again use the harmonic oscillator 
approximation 
for $\hat h_{{\rm
rel},i}$ given before
\ba \hat h_{s,i}|\vec k_i\rangle&=&\frac{\vec P_{s_i}^2}{2^{5/2} 
x_i}|\vec k_i\rangle
=\frac{\vec k_i^2}{2^{5/2}x_i}|\vec k_i\rangle\nonumber\\
\hat h_{{\rm rel},i}|n_{r,\ell}\rangle&=&\left(
\frac{\vec p_{i}}{m_{\rm red} (i)}
+\kappa_{\rm eff} r_i-
2\sqrt\kappa_{\rm eff}\right)|n_{r,\ell}\rangle\nonumber\\
&=&\left(\alpha_{n_r,\ell,i}\sqrt[3]{\frac{\kappa^2_{\rm 
eff}}{m_{\rm 
red}(i)}} -2\sqrt\kappa_{\rm eff}\right)|n_{r,\ell}\rangle\ea
and 
\be \alpha_{n_r,e,i}=(2n_r+\ell+2.3381).\ee
We consider the integration over the auxiliary parameters $x_i$ 
and $y_i$ with the
measure $\frac{1}{\sqrt{x^2_i-y^2_i}}$ as a summation over 
intrinsic degrees of
freedom of the meson, i.e.
\be \frac{\sqrt2}{\pi}\beta\int^\infty_0
dx\int^x_{-x}\frac{dy}{\sqrt{x^2-y^2}}=\sum_x\sum_y,\ee
then each meson is characterized by the intrinsic parameters $| 
r\rangle=|k_r,\omega_{n_r},
\ell,x,y\rangle$. The meson  many-body wave function is a 
symmetrized state defined
by the occupation numbers of each state $|r\rangle$.
The free energy of the mesons has the usual Bose gas form, but 
now including
the summation over the $x_i$ and $y_i$ coordinates
\ba F&=&\frac{1}{\beta}\sum_{f \bar
f}\sum_n\sum_k\sum_x\sum_y\ln(1-\exp[-
\beta\varepsilon(n,k,x,y)])\nonumber\\
&=&\sum_{f \bar f}\sum_n g(n)\frac{\sqrt2}{\pi}\int^\infty_0 
dx\int^x_{-x}
\frac{dy}{\sqrt{x^2-y^2}}\frac{V}{(2\pi)^3}\int d^3
k\ln(1-\exp(-\beta\varepsilon))\nonumber\\
&=&\sum_{f\bar f}\sum_n 2^{5/4}g(n)\frac{\beta^{-1.5} 
V}{\pi^{5/2}}\int dx
x^{3/2}\int\frac{dy}{\sqrt{x^2-y^2}}
\sum^\infty_{s=1}\frac{1}{s^{5/2}}\exp[-s\beta\varepsilon_r]\ea
where \be\varepsilon 
(n,k,x,y)=\frac{k^2}{2^{5/2}x}+\varepsilon_r\ee
and \be\varepsilon_r=\omega_n+\sqrt2 x+\frac{m^2}{2^{3/2}(x-
y)}+\frac{m^2}
{2^{3/2}(x+y)}.\ee
We follow the standard way to evaluate the free energy.
We first
integrate by parts and then rewrite the denominator as a power 
series before finally
integrating the energy of relative motion  numerically.
To complete 
the discussion
of the meson gas, we also give the expressions for the
entropy and energy density
\ba p&=&-F/V;\nonumber\\ 
s&=&-\frac{1}{V}\left(\frac{\partial F}{\partial 
T}\right)_V\nonumber\\
&=&\frac{3}{2}\frac{p}{T}+\sum_{f \bar f}\sum_n g(n)
2^{1/4}\frac{\beta^{1/2}}{\pi^{5/2}}\int dx x^{3/2}\int
\frac{dy}{\sqrt{x^2-y^2}}\nonumber\\
&&\left[\varepsilon_r-\frac{1}{\beta} \frac{\partial \varepsilon_r}
{\partial  T}
\right]\sum^\infty_{s=1}\frac{\exp(-\beta 
s\varepsilon_r)}{s^{3/2}};\nonumber\\
u&=&T s-p.\ea

The resulting pressure of the meson gas is shown in fig.~2
for two cases (a) 
for the ground
state mesons only  and (b) for the lowest $N=15$
states.
\begin{figure}[hbt]
\unitlength1cm
\begin{center}
\begin{picture}(15,9)
\epsfbox{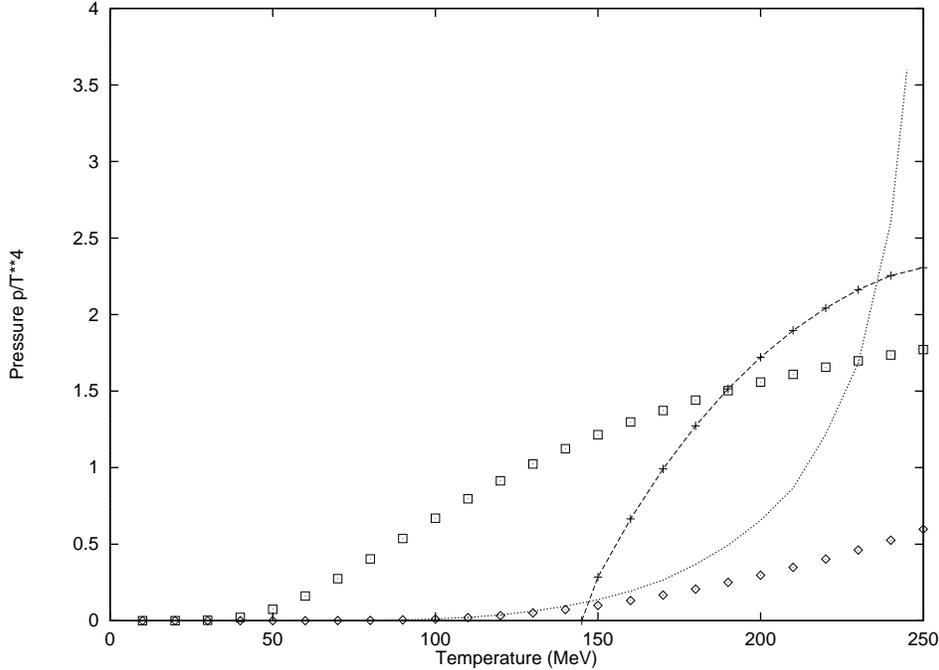}
\end{picture}
\end{center}
\caption{\label{afig2} Normalized pressure $p/T^4$ as a function of
temperature $T$. The lowest pressure  is obtained 
for  the confined mesons when only the ground state mesons 
are considered. A diverging pressure (drawn line)
results when mesons with $N=15$ main quantum numbers are taken into account.
For comparison the free $q \bar q$ gas pressure (open boxes) and
the correlated $q \bar q$ pressure (+) are shown.}
\end{figure}
One sees that because of the large meson masses the pressure 
is rather
small up to $T=150$ MeV. Around this temperature  the large 
number of degrees 
of freedom for $N=15$  makes a 
difference. At a temperature $T=250$ MeV a Hagedorn transition 
occurs, when  the number of
relevant states and consequently the pressure increase 
tremendously. 
One sees that the treatment of the system as a gas of noninteracting
mesons becomes
unrealistic at high temperatures, since
the mean distance $r_d$ between mesons  
becomes smaller than the size of the mesons.
The Bose
distribution yields the mean particle number $\langle N\rangle$  
\be\langle N\rangle=V\sum_{f \bar 
f}\sum_n(2^{5/4})g(n)\frac{\beta^{-1.5}}{\pi^{5/2}}
\int dx x^{3/2}\int
dy\frac{1}{\sqrt{x^2-y^2}}\sum^\infty_{s=1}\frac{1}{s^{3/2}}
\exp[-s\beta\varepsilon_r]\ee
and
\be r_d=\left(\frac{1}{\frac{4\pi}{3}\frac{\langle 
N\rangle}{V}}\right)^{1/3}.\ee
The function $r_d(T)$ in  fig. 3 is a strongly decreasing function. At 
$T\simeq 100$
MeV the mean distance between mesons is comparable
to the root mean square radius $\sqrt{\langle 
r^2\rangle}$ of the ground
state meson
\be \sqrt{\langle 
r^2\rangle}=\left(\frac{2.33}{\sqrt[3]{\kappa_{\rm
eff}^2(T)/m}\cdot m}\right)^{1/2}.\ee
Around  this temperature the approximation of a quasifree meson gas 
fails, the mesons
start to overlap and interact strongly. 
We also remark that the expression for the entropy leads to a divergent 
term at
$T=T_c=260$ MeV. This is irrelevant because the assumption
of
independent mesons already
fails at much lower temperatures as discussed 
above.
Clearly, for higher
temperatures it becomes more and more
inefficient to treat the system as a meson gas and a calculation in 
terms of quark
variables is more advantageous. 
\begin{figure}[hbt]
\unitlength1cm
\begin{center}
\begin{picture}(15,9)
\epsfbox{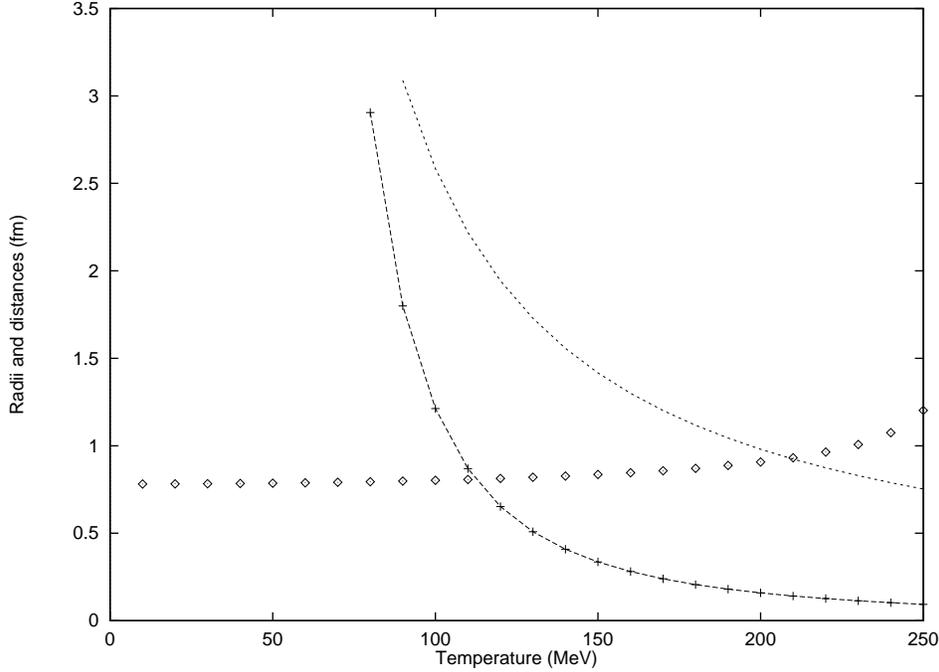}
\end{picture}
\end{center}
\caption{\label{afig3} 
The meson groundstate rms radius (squares) as a function of temperature and
the mean distances between mesons $r_d$  (+)
and between quarks and antiquarks 
of opposite color $R_{qc}$ (dashed line).}
\end{figure}
This does not mean that the strings 
and their
attractive forces have disappeared. The strings will
arrange themselves according to the
quark antiquark distributions, instead of
extending 
between fixed pairs of
quarks and antiquarks. In fig. 3 one can estimate
that near $T = 140$ MeV the mean distance $R_{qc}$
between quarks and antiquarks of opposite color
is equal to twice the rms meson radius.
After this energy the correlated $q \bar q$ system
starts to be competitive in energy with the meson system.
This will be treated in the next section. 

\section{Correlated quarks and antiquarks at high temperatures}

When the mean distance between quarks and antiquarks becomes 
smaller than twice the rms
radius of the meson state, the mesons will no longer keep their 
identity, i.e. the
strings will rearrange themselves in configurations with the 
smallest free
energy. An exact calculation probably can only be done with a 
Monte Carlo
simulation. Here we would like to present an estimate using the 
Hamiltonian given before. The calculation is based on the nearest
neighbour saturation of color forces, a concept which has been used
in the many quark problem before \cite {RBS}.
In the high temperature limit we consider 
${\cal H}$
as a sum of noninteracting quark and antiquark energies ${\cal 
H_0}$
and a perturbation ${\cal H_I}$
\ba
{\cal H}&=&{\cal H}_0+{\cal H}_I\nonumber\\
{\cal H}_0&=&\sum_{q}\sqrt{\vec p^2_q+m^2}+\sum_{\bar q}\sqrt{\vec 
p^2_{\bar q}+m^2}\ea
\be {\cal H}_I=\sum_{q\bar q\atop ff'}(\kappa_{\rm eff}|\vec r_q-\vec
r_{\bar q}|-2\sqrt\kappa_{\rm eff}).\ee  
Here the summation over $r_q,r_{\bar q}$ distances is understood 
under the
constraint of a minimal total length of strings. In thermal 
perturbation theory we
first evaluate the free energy related to ${\cal H}_0$
\ba F_0&=&-\frac{1}{\beta}\ln Z_{\bar q q}=\sum
\ln(1+e^{-\beta\varepsilon_p})\nonumber\\
&=&-\frac{24}{\beta}V\int\frac{d^3 p}{(2\pi)^3}\ln(1+e^{-
\beta\sqrt{\vec
p^2+m^2}}).\ea
The mean number of quarks with a color $c$ 
equals the mean number of antiquarks with opposite color $\bar c $
\ba\langle N_{qc}\rangle&=&\langle N_{\bar 
q \bar c}\rangle= 4 V\int\frac{d^3p}{(2\pi)^3}
\frac{1}{e^{\beta\varepsilon_p}+1};\nonumber\\
&=&4 V\sum^\infty_{s=1}(-1)^{s+1}\int\frac{d^3p}{(2\pi)^3}
e^{-\beta\varepsilon_ps}\nonumber\\
&=&4 V \frac{1}{2\pi^2}m^3\sum^\infty_{s=1}(-1)^{s+1}
\left(\frac{2}{\beta^2s^2m^2}K_1(\beta sm)
+\frac{1}{\beta sm}K_0(\beta sm)\right).\ea
We imagine that each quark is localized in a Wigner Seitz cell of 
radius $R_{q c}$. The numerical values of $R_{q c}$ as a function
of temperature are shown in fig. 3.
The quasilattice of Wigner size cells also  of radius $R_{q c}$ for the 
antiquarks 
may be shifted
relative to the quarks by a distance $|\vec \ell|< 2R_{qc}$
\be R_{q c}=\left(\frac{1}{\frac{4\pi}{3}\frac{\langle 
N_{qc}\rangle}{V}}\right)^{1/3}.\ee
Then we can 
calculate a
distribution function for a given quark of color $c$
to find the
nearest antiquark with color $\bar c $ at a distance $r$ away.
\ba g(r)&=&\frac{1}{\left(\frac{4\pi}{3}R_{q c}\right)^3}\int d\vec 
r_q \ \theta(R_{q c}-|\vec
r_q|)\nonumber\\
&&\qquad \int d\vec r_{\bar q} \ \theta(R_{q c}-|\vec\ell+\vec 
r_q|)\nonumber\\
&&\qquad \int d\vec\ell \ \theta(2R_{q c}-|\vec\ell|)\delta(\vec r-
(\vec 
r_q-\vec r_{\bar
q})).\ea
In fig. 4 we give the scaled distribution function 
$g(r/R_{q c})(r/R_{q c})^2$ and compare
it to $\theta(1-r/R_{q c})(r/R_{q c })^2$.
\begin{figure}[hbt]
\unitlength1cm
\begin{center}
\begin{picture}(15,9)
\epsfbox{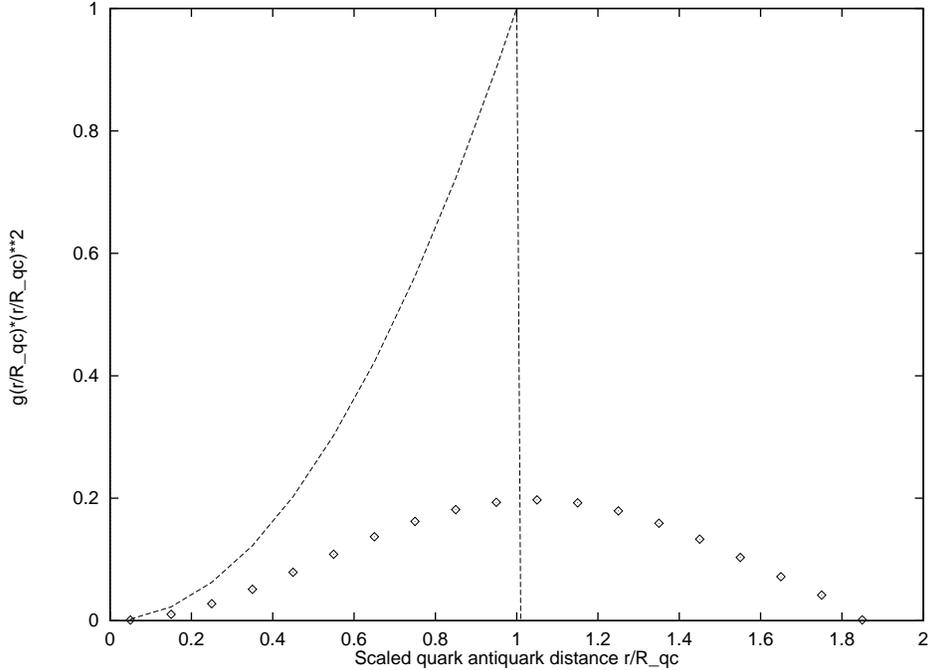}
\end{picture}
\end{center}
\caption{\label{afig4} The distribution function $g(r/R_{qc})(r/R_{qc})^2$
(squares) is  compared to the uncorrelated theta function multiplied by
phase space (line).}
\end{figure}

These two curves have approximately the maximum at the same 
place, but the
correlation function extends  to larger distances. With this 
numerically calculated
function we calculate the modified free energy in perturbation 
theory
\be F=F_0+ 3 N_{qc} N_{\bar 
q \bar c }\langle {\cal
H}_I\rangle \frac{V_{\rm corr}}{V}\ee
where $\langle {\cal
H}_I\rangle\cdot V_{\rm corr} =4\pi\int g(r/R_{qc})r^2 dr(
\kappa_{\rm eff} r -
2\sqrt\kappa_{\rm eff})$. The
correlation volume $V_{\rm corr}=4\pi\cdot R^3_{qc}\cdot 0.203$ is 
smaller than the
naive volume and the mean $\langle r\rangle=\int r^3 dr 
g(r/R_{qc})/\int r^2 
dr
g(r/R_{q c})$  is unessentially larger than $R_{q c}$
\be \langle r\rangle=1.03 R_{qc}.\ee
In fig.2  we show the pressure of the correlated quark 
antiquark gas  times
$1/T^4$ (with symbols $+$). It rises rapidly around $T=150$ MeV 
and overshoots the 
uncorrelated pressure at
$T\approx 175$ MeV. Near $T=145$ MeV it already 
exceeds the still very
small pressure  of the meson gas. The meson contribution
is damped by 
roughly  $e^{-2m/T}$ 
compared to the quark excitations which have Boltzmann 
weights $e^{-m/T}$.
The average interaction potential changes sign at $T=175$ MeV. Of 
course this value
depends critically on the subtraction $2\sqrt{\kappa_{\rm eff}}$ in 
the 
$q \bar q $ potential, which gives
good agreement with the rho-meson in our model. The inclusion of
 spin-dependent
interactions may modify the 
potential.  The
rather simplified calculation presented here conveys 
the correct message. It is not 
the
``deconfinement''  temperature $T_c=260$ MeV for pure glue 
theory which determines the 
phase transition in the presence of quarks, but the
lower temperature $T\simeq 145$  MeV where the quarks and 
antiquarks liberate themselves from the fixed meson configurations.
At and above this temperature the quarks and antiquarks are still
correlated by  strings with a string tension of 
$\kappa^{1/2}_{\rm eff}(T=145$
MeV)=400 MeV, but these strings change their partners 
constantly.  If this picture is correct, the heavy quarkonia
would still experience a long range potential in addition
to the modified Coulomb potential due to screening.
Also a heavy quark would find additional
light antiquarks nearby at no cost in energy. It is in this sense
that heavy quarkonia dissolve. The  Hagedorn transition of 
the resonance gas
occurring  at higher temperature is irrelevant, since it happens
outside of the range
of credibility of the model. It would actually prefer a system 
of meson
re\-sonances instead of the quark antiquark gas,
but at this temperature the resonances are
strongly interacting and the independent 
meson model has lost
all its theoretical foundation.

\section{Chiral dynamics of constituent quarks}

Having set up the formalism for a fixed constituent quark mass 
we are ready to
consider the variation of the constituent quark mass due to the 
quark condensate.
We can use  the linear $\sigma$-model to generate the quark mass 
$m=g\langle
\sigma\rangle$ where $\langle \sigma\rangle$ is determined from 
the minimum of the
free energies either in the resonance gas phase or in the correlated 
quark antiquark
phase plus the potential energy of the mean $\sigma$-field
\ba \hat F_{RG}(\langle\sigma\rangle)&=&\sum_{f\bar f}\sum_n 
2^{5/4}
g(n)\beta^{-3/2} V\nonumber\\
&&\int^{\infty}_{0} dx x^{3/2}\int^{+x}_{-x} 
dy\frac{1}{\sqrt{x^2-y^2}}
\sum^\infty_{s=1}\frac{1}{s^{5/2}}e^{-
s\beta\varepsilon_r}\nonumber\\
&&+U(\langle\sigma\rangle)\ea
where
\ba 
\varepsilon_r&=&(2n_r+\ell+2.34)\sqrt[3]{\frac{\kappa^2_{\rm
eff}(T)}{\sqrt2(x-y^2/x)}}- 2 \sqrt\kappa_{eff(T)}\nonumber\\
&&+\sqrt2
x+\frac{(g\langle\sigma\rangle)^2}{2^{3/2}(x-
y)}+\frac{(g\langle\sigma\rangle)^2}
{2^{3/2}(x+y)}.\ea
Here one sees how the relativistic bound state calculation takes
into account
the mean field
masses $g\langle\sigma\rangle$ 
of the quarks via the integration over the auxiliary 
parameters $x$ and $y$.

In the high temperature phase of correlated quarks and antiquarks 
with variable
masses the free energy reads
\ba \hat F_{CQ}^{(\langle\sigma\rangle)}&=&-\frac{24}{\beta} 
V\int\frac{d^3\vec p}
{(2\pi)^3}\ln(1+\exp(-\beta\sqrt{\vec 
p^2+(g\langle\sigma\rangle)^2}
))\nonumber\\
&&+3 N_{qc}(g\langle\sigma\rangle)N_{\bar 
q \bar c}(g\langle\sigma\rangle)\langle{\cal
H}_I\rangle \frac{V_{\rm corr}}{V}\ea
where also the interaction energy and the correlation volume 
depend on the 
constituent quark mass $m=g\langle\sigma\rangle$, as described 
in section IV.
The mean field quark mass is determined by the condition that the 
free energies
$\hat F(\langle\sigma\rangle)$ are minimal, i.e.
\ba \frac{\partial\hat
F}{\partial\langle\sigma\rangle}\Bigm\vert_{\langle\sigma\rangle
=\sigma_0}&=&0;
\nonumber\\
m&=&g\sigma_0.\ea
In fig.5  we give the dependence of the constituent quark mass on 
the 
temperature for
quark matter and resonance matter. 
\begin{figure}[hbt]
\unitlength1cm
\begin{center}
\begin{picture}(15,9)
\epsfbox{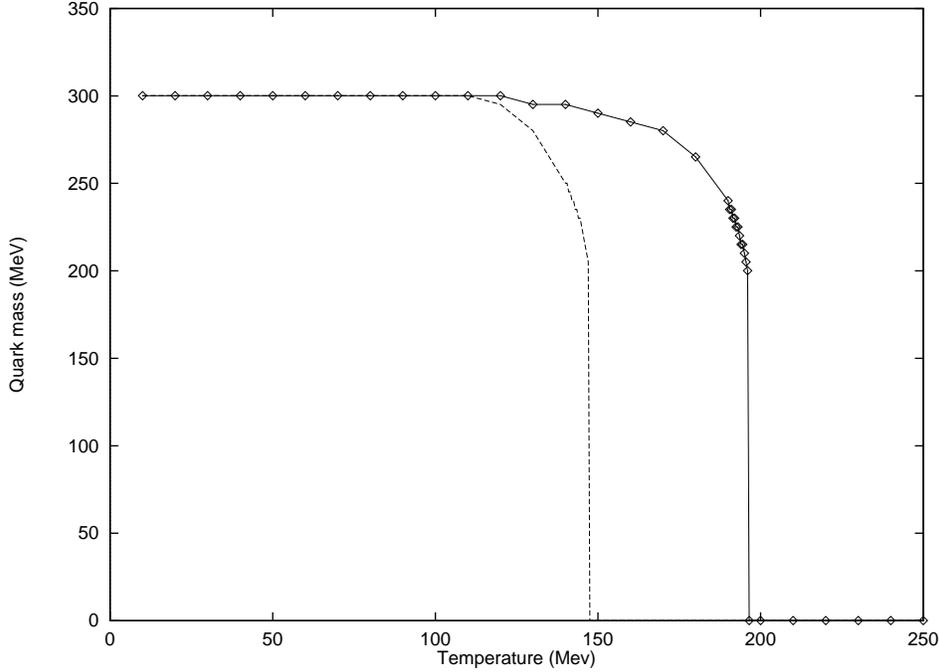}
\end{picture}
\end{center}
\caption{\label{afig5} The constituent quark mass 
as a function of the temperature is shown for the
correlated quark antiquark (lines) and for the resonance gas
(squares).}
\end{figure}
In both cases we have a first 
order phase
transition with $T_{\rm chiral}(CQ)=147$ MeV, $T_{\rm 
chiral}(RG)=196$ MeV.
Until 120 MeV very little  change occurs. This is different from
calculations of the
NJL type with  nonconfining interactions.  In the line of 
argumentation  given
before we do  not attribute great significance to the higher 
chiral phase
transition of resonance meson matter, probably a more accurate 
calculation including the low lying $\pi$- and $\sigma$-degrees
of freedom explicitly
would make the two chiral transitions 
coincide.  What is very interesting 
is the almost coincidence between the deconfinement temperature
$T_{\rm dec}=143$ MeV, 
where the resonating
meson gas goes over into correlated quark matter and $T_{\rm 
chiral}(CQ)=147$ MeV.
This coincidence is surprising given the simplicity of the
theoretical calculation. It is dependent on the 
choice of original constituent quark mass and the 
behavior of the string tension with temperature.
It may indicate that the constituent quark mass
is intimately related to the gluon dynamics and 
the choice of critical behavior of the string tension is perhaps
not so far off from reality.

\section {Conclusions}

In the framework of the chiral constituent model we have presented a 
relativistic calculation of the pressure and the effective quark mass 
of quarks and antiquarks at finite temperature. At low temperature the 
constituents are bound in mesons which change their composition only very slowly.
Due to the decreasing string tension the quark antiquark bound states 
have a slightly decreasing mass. In future work we would like to 
investigate further the $\rho$ meson spectrum as a function of temperature.
At temperatures around $T=120$ MeV the approximation of noninteracting mesons 
becomes inaccurate and a description in terms of liberated quarks and
antiquarks connected by varying string configurations becomes more appropriate.
At a temperature $T\approx 145$ MeV both a deconfinement and 
chiral transition occurs
to massless fermions which prefer arranging their string to the nearest
antiquark of opposite color nearby. 
This temperature is distinctly lower than the quark mass 
transition in  the resonance gas
occuring at $196$ MeV.
Important fluctuations of the chiral fields are 
still missing in the calculation so far. 
Especially the Goldstone $\vec\pi$-fields 
contribute an important part of the pressure at
low temperatures. The 
addition of these extra
degrees of freedom will be done in another paper. In the 
resonance gas the
summation over $q\bar q$-bound states should exclude the  $^1 
S_0(\vec\pi)$ and $^3
P_0{(\sigma)}$ states. Therefore, for the   ground state $^3 S_1$ 
and $^1 S_0$
state the new degeneracy factor becomes   $ g'(0)=16-3=13$ and 
$g'(1)=48-1=47$ for the 
first excited
states, where the 48 states come from the 
degenerate 4
isospin $^3 P_0,^3 P_1,^3 P_2$ and $^1P_1$ channels in the harmonic 
oscillator
model. 
To evaluate the fermion determinant in the presence of $\sigma$ and
$\vec \pi $ fields it is advantageous to use the heat kernel method
which has been developed for finite temperatures in ref. \cite {S}.
The methods described in this article are not limited to finite
temperature. They can as well be applied to finite baryon density,
especially to the nuclear problem which has the most promising 
future. Until now it has resisted any explanation in
terms of fundamental quark and string degrees of freedom. Constituent
quarks, however, have the correct dynamics in terms of mesonic chiral
interactions and confining string dynamics to make the problem
accessible.

\section*{Appendix}
\setcounter{equation}{0}
\renewcommand{\theequation}{A.\arabic{equation}}

The density matrix for the relativistic two body scalar problem 
with interaction $V$ neglecting retardation and backward motion has 
the form:

\be
\rho_{q\bar q}=\exp[-\beta( \sqrt{\vec p_q^2 +m^2}+
\sqrt{\vec p_{\bar q}^2 +m^2} +V(|\vec r_{\bar q}-\vec r_q|))].
\ee

It has a path integral representation

\be
\rho_{q\bar q}=\int \cal D \vec r_{q}  \cal D \vec r_{\bar q}
\exp[-\int ^{\beta}_0[m \sqrt{1+\dot{\vec r_q}^2} +
m \sqrt{1+\dot{\vec r_{\bar q}}^2}  +V(|\vec r_{\bar q}-\vec r_q|)]d\tau].
\ee

This integral can be divided up into a product of integrals 
over small $\Delta \tau$  intervals which are reformulated
with auxiliary fields. The integrals with square roots 
are transformed into integrals without square roots with the help of
the following integral:

\be\label{xx}
\exp{-\sqrt a }=\frac{2}{\sqrt \pi}\int dx \exp{-(x^2+\frac{a}{4 x^2})}
\ee
Eq. (\ref{xx}) can be proven by  evaluating the integral $I$ and its derivative
with respect to $a$
\be
I=\int dx \exp [-(x-\frac{\sqrt a}{2 x} )^2],
\ee
\be
\frac{d I}{d a}=0.
\ee
Since the derivative of $I$ with respect to $a$ is zero,
the value of $I$ is the norm of the Gaussian integral
$I(a=0)$. After a rescaling $x^2=\mu_1^2 \Delta \tau$ and a suitable choice of
the measure we obtain the form
\be
\exp[-m \sqrt{1+\dot{\vec r_q}^2}\Delta \tau] = \cal N \int d\mu_1 
\exp[-(\mu_1^2+\frac{m^2}{4 \mu_1^2}(\dot{\vec r_q}^2+1)) \Delta \tau]
\ee
Multiplying the individual time slices with each other one generates the
full density matrix
\ba
\rho_{q\bar q}&=&\cal N^{'} \int \cal D \vec r_{q}  \cal D \vec r_{\bar q}
\cal D \mu_1 \cal D \mu_2\\
&&\quad \exp[-\int ^{\beta}_0
[\mu_1^2+ \mu_2^2+\frac{m^2}{4 \mu_1^2} (1+\dot{\vec r_q}^2) +
\frac {m^2}{4 \mu_2^2} (1+\dot{\vec r_{\bar q}}^2)  
+V(|\vec r_{\bar q}-\vec r_q|)]d\tau.\nonumber
\ea
In general the auxiliary fields are functions of $\tau$.
We propose as an approximation  to keep only auxiliary fields  
in the functional integral which are independent of time.
Then the  path integral over $\cal D \vec r_{q}  \cal D \vec r_{\bar q}$ can be
executed by solving the corresponding two-body Schroedinger equation
\be
[-\frac{{\vec \nabla_q}^2}{4 \mu_1^2}-\frac{{\vec \nabla_{\bar q}}^2}{4 \mu_2^2}
+V(|\vec r_{\bar q}-\vec r_q|)] \Psi(\vec r_{q},\vec r_{\bar q})=
E(\mu_1,\mu_2)\Psi(\vec r_{q},\vec r_{\bar q}).
\ee
The final integration over the auxiliary fields becomes a two-dimensional
integration
\be
\rho_{q\bar q}=\cal N^{'} \int 
d \mu_1 d \mu_2 \exp[-\int ^{\beta}_0
[\mu_1^2+\mu_2^2+ \frac{m^2}{4 \mu_1^2} +
\frac {m^2}{4 \mu_2^2} 
+E(\mu_1,\mu_2)]d\tau].
\ee
Finally the partition function can be written in the form of eq. $8$ 
of section $1$ by the
substitution $\beta \mu_1^2\rightarrow \mu_1^2$ and
$\beta \mu_2^2\rightarrow \mu_2^2$.  The normalization factor in 
front of the integral in eq. $8$ 
is adjusted to reproduce the free partition function without potential.

\section*{Acknowledgements}
We would like to thank B. J. Schaefer for
a careful reading of the manuscript and
our colleagues in the Rhein-Main-Neckar group on Relativistic
Heavy Ion Physics
for their critical and
helpful comments.

\end{document}